\documentclass[10pt,prl,aps,showpacs,twocolumn,unsortedaddress]{revtex4-1}
\usepackage{graphicx,bm}
\usepackage{amsmath}
\usepackage{amssymb}
\usepackage{epsfig}
\usepackage{epsf}
\usepackage{verbatim}
\usepackage{hyperref}
\usepackage[usenames]{color}

\newcommand{\dd}{\, \mathrm{d}}

\newcommand{\ve}{\varepsilon}

\newcommand{\oo}{\Omega}
\newcommand{\bvk}{\boldsymbol{k}}
\newcommand{\bs}{\boldsymbol}
\newcommand{\sk}{\text{sk}}

\def\be#1\ee{\begin{equation}#1\end{equation}}
\def\ba#1\ea{\begin{align}#1\end{align}}
\def\bal#1\eal{\begin{equation}#1\end{equation}}

\newcommand{\ua}{\uparrow}
\newcommand{\da}{\downarrow}
\begin{document}

\title{
Magnetization relaxation and geometric forces in a Bose ferromagnet
}

\author{J. Armaitis}
\email{j.armaitis@uu.nl}
\author{H.T.C. Stoof}
\author{R.A. Duine}
\affiliation{Institute for Theoretical Physics, Utrecht
University, Leuvenlaan 4, 3584 CE Utrecht, The Netherlands}

\date{\today}

\begin{abstract}
We construct the hydrodynamic theory for spin-1/2 Bose gases at arbitrary temperatures. This
theory describes
the coupling between the magnetization, and the normal and superfluid components of the gas.
In particular, our theory contains the geometric forces on the particles that arise from
their spin's adiabatic following of the magnetization texture.
The phenomenological parameters of the hydrodynamic theory are calculated in the Bogoliubov approximation 
and using the Boltzmann equation in the relaxation-time approximation. We consider the topological Hall 
effect due to the presence of a skyrmion, and show that this effect manifests itself in the collective modes of 
the system. The dissipative coupling between the magnetization and the normal component is shown to give 
rise to magnetization relaxation that is fourth order in spatial gradients of the magnetization direction.
\end{abstract}

\maketitle

\textit{Introduction.---} Geometric forces are abundant in virtually all areas of physics,
from the classical conical pendulum \cite{sakurai} to a single quantum spin \cite{PhysRevLett.97.190401}.
In particular after the advent of the Berry phase \cite{Berry08031984}, a large number of 
manifestations of geometric forces, including the 
optical Magnus effect \cite{Bliokh2008},
the topological Hall effect \cite{PhysRevLett.102.186602, PhysRevLett.110.117202}
and geometric forces due to synthetic gauge fields \cite{RevModPhys.83.1523} have been 
predicted and observed. 

In metallic ferromagnets, magnetization dynamics leads to forces on quasiparticles of geometric origin called 
spin motive forces, that have gained considerable attention recently 
\cite{PhysRevLett.98.246601,PhysRevB.77.014409,PhysRevB.79.014402,
PhysRevLett.102.067201,PhysRevLett.108.147202}. 
Furthermore, spin textures with nonzero chirality, such as the skyrmion lattice observed
recently \cite{Mühlbauer13022009,Yu2011} induce the so-called topological Hall effect \cite{PhysRevLett.102.186602, PhysRevLett.110.117202}.
In addition, the coupling between magnetization and quasiparticles has also been shown to give rise to novel forms of magnetization relaxation
in this case. A prominent example is inhomogeneous Gilbert damping  \cite{PhysRevB.78.140402,PhysRevLett.102.086601,PhysRevB.81.100403}. This effect is important in clean solid-state systems.
We therefore expect these effects to be particularly important for gases of
ultracold atoms, that, in contrast to conventional condensed-matter 
systems, are free of impurities.

The field of ultracold atoms is characterised by exquisite
experimental control \cite{Anderson14071995,PhysRevLett.75.3969, RevModPhys.82.1225}.
Relevant for our focus is the great amount of recent activity on spinor Bose gases. Firstly, it has been discovered that
these gases can be either ferromagnetic or antiferromagnetic depending on the details of the scattering lengths 
\cite{PhysRevLett.81.742,Machida}. Furthermore, numerous studies at zero temperature, based on the 
Gross-Pitaevskii equation, have elucidated the long-wavelength properties of spinor gases
\cite{PhysRevA.77.063622, PhysRevB.80.024420}. Other areas of current interest in the field include
topological excitations, magnetic dipole-dipole interactions and non-equilibrium quantum dynamics \cite{Kawaguchi2012253}. The recent progress
in the understanding of ferromagnetic spinor gases and their manipulation by light has also enabled detailed studies of 
magnetization dynamics \cite{Sadler2006,2012arXiv1205.1888S}.

Despite these activities, a theory that describes simultaneously all phases of ferromagnetic spinor Bose gases
and includes 
both geometric and dissipative coupling between superfluid, ferromagnetic order parameter
and quasiparticles
 is lacking. 
The purpose of this Letter is to put forward such a theory. 
This theory is needed to determine properties of collective modes  at
arbitrary temperatures that consist of combined dynamics of the ferromagnetic and
superfluid order with the normal component of the gas. These results can e.g.\ be
used to detect the presence of skyrmions and their dynamics in the gas. We
choose to work in the hydrodynamic regime, where the
coupling between the magnetization, and the normal and superfluid components of the
gas is controlled by a gradient expansion. This approach is valid in the regime where local equilibrium is 
enforced by frequent collisions.
In addition, we make a connection between long-wavelength and 
microscopic physics by calculating all the hydrodynamic parameters from first principles.
In the microscopic determination of the parameters, we focus on the spin-1/2 case leaving higher spin for future work.

\textit{Hydrodynamic equations.---} Describing a ferromagnetic Bose gas requires considering three phases: 
unpolarized normal fluid, normal ferromagnet and superfluid ferromagnet. These three phases have different 
sets of relevant hydrodynamic variables, that have to be taken into account in order to fully describe the behavior of the 
system. A complete set of relevant variables includes the order parameters and the 
conserved quantities. The order parameters of a homogeneous Bose ferromagnet are the superfluid velocity 
$\bs v_s$ and 
the magnetization density $P n \oo^\alpha$. 
Here $P$ is the polarization, and $\oo^\alpha$ is the 
dimensionless magnetization direction, normalized such that $\oo^\alpha\oo^\alpha=1$. (We use the Einstein summation convention throughout the paper.)
The conserved quantities are the total particle density 
$n$, the total particle
current $\bs j$, the magnetization density $P n \oo^\alpha$ and the energy. In the hydrodynamic 
approach we write for each conserved quantity a continuity equation. Not considering energy conservation
for simplicity, we therefore have the following set of hydrodynamic equations:
\ba
\label{density}
\partial_t n + \bs \nabla \cdot \bs j = 0,\\
\label{spin}
\partial_t \left( P n \oo^\alpha \right)
+{ \bs \nabla } \cdot \left( (\bs j_\mathrm{spin}^\perp)^\alpha + (\bs j_\mathrm{spin}^\parallel)^\alpha
\right)
= 0
,\\
\label{totmom} 
m\partial_t \bs j 
+ m \bs \nabla \cdot \bs \Pi
+ n \bs \nabla V
- P n \bs E
- P \bs j \times \bs B  = 0.
\ea
In these equations, $m$ is the particle mass,
$\bs \Pi = n_n \bs v_{n} \bs v_{n} + n_s \bs v_{s} \bs v_{s} + \bs 1\, p/m$ is the energy-momentum
tensor, and $V$ is the trapping potential. We have also introduced the pressure $p$, the normal fluid
velocity $\bs v_n$ and the normal fluid density $n_n$, and equivalent quantities $\bs v_s$ and 
$n_s$ for the superfluid. We use these quantities to define the normal and superfluid particle currents
$\bs j_n = n_n \bs v_n$ and $\bs j_s = n_s \bs v_s$, such that $\bs j = \bs j_n + \bs j_s$. The total density is then $n = n_s + n_n$.
Note that coordinate space tensors and vectors are 
denoted by bold font, while spin space vector 
components are denoted by Greek superscripts.

The coupling between magnetization and normal fluid leads to geometric forces
\cite{0022-3719-20-7-003,PhysRevLett.68.1022}.
These can e.g.\ be understood as resulting from the Berry curvature and spin Berry phases that
the atoms pick up as their spin adiabatically follows the magnetization texture.
More concretely, there exist now an artificial electric field
\be
\bs E
=
\hbar \ve^{\alpha \beta \gamma} 
\oo^\alpha (\partial_t \oo^\beta) (\bs \nabla \oo^\gamma)/2
\ee
and an artificial magnetic field
\be
\bs B
=
- \hbar \ve^{\alpha \beta \gamma} 
\oo^\alpha (\bs \nabla \oo^\beta) \times (\bs \nabla \oo^\gamma)/4,
\ee
where we have introduced the totally
antisymmetric Levi-Civita tensor $\ve^{\alpha \beta \gamma}$. As we have already seen in Eq.\ \eqref{totmom}, 
these fields enter the hydrodynamic equations as the electric and magnetic parts of the Lorentz force, respectively, 
acting on the particle current.

A more detailed discussion on the spin currents is now in order. The longitudinal spin current describes spin
transport with spin polarization along the magnetization direction due to particle currents: $( \bs j_\mathrm{spin}^\parallel)^\alpha = P \bs j \oo^\alpha$. 
The transverse spin current has, to lowest order in the gradient expansion, terms 
proportional to the spin stiffness $A_\mathrm{s}$ and a parameter proportional to the transverse 
spin diffusion constant $\eta_\perp$ \cite{PhysRevB.79.094415}:
\ba
(\bs j_\mathrm{spin}^\perp)^\alpha = 
- \ve^{\alpha \beta \gamma} \frac{1}{n} \eta_\perp \oo^\beta  
(n \partial_t &+ \bs j \cdot \bs \nabla) \bs \nabla \oo^\gamma 
\qquad \qquad \nonumber \\
&- A_s \ve^{\alpha \beta \gamma} \oo^\beta \bs \nabla \oo^\gamma
.
\ea
The first term describes transverse spin relaxation. It contains the usual hydrodynamic derivative
and illustrates the fact that only gradients of magnetization relax because
spin is conserved in our system. The second term
represents non-dissipative transverse spin transport. Note that the above results can be understood
as containing all symmetry-allowed terms up to second order in gradients.
The superfluid velocity obeys the Josephson relation
\be
\label{sfmom} 
m\partial_t \bs v_s
+ \bs \nabla \left[
m v_s^2/2 + V+\mu
\right] 
- \bs E = 0,
\ee
where $\mu$ is the chemical potential, and also the Mermin-Ho relation
\be
\label{merminho}
m\bs \nabla \times \bs v_s = -\bs B.
\ee
This completes the set of hydrodynamic equations.

Our hydrodynamic equations correctly describe the system at any
temperature relevant for cold-atom experiments. At sufficiently low temperatures both order parameters are non-zero, and we have
a ferromagnetic Bose-Einstein condensate with a negligible normal fluid density. Setting the 
normal fluid density and its velocity to zero, and setting the polarization to $P=1$, 
we obtain the well-known $T=0$ limit \cite{Kawaguchi2012253}. At higher 
temperatures, where the normal fluid density is sizable, we have to use the full set of equations. 
Heating the system even further results in the disappearance of the Bose-Einstein condensate at 
$T_\text{BEC}$. (Note that in general the critical temperature for Bose-Einstein condensation is lower 
than the ferromagnetic transition temperature, as shown in Ref.\ \cite{TfmBiggerTbec}.) 
We then have to discard Eqs.\ \eqref{sfmom} and \eqref{merminho}, and set 
the superfluid density as well as its velocity to zero. Finally, in the high temperature limit 
$T > T_\text{FM}$, the average magnetization is zero, and the distinction between longitudinal and 
transverse spin polarization disappears. In addition, there are no artificial electromagnetic fields and 
consequently the geometric forces vanish. For those reasons, the spin current becomes
$\bs j_\mathrm{spin}^\alpha = - D_s \bs \nabla (n \oo^\alpha)$, where $D_s$ is the spin diffusion constant,
which is related to the longitudinal spin relaxation time determined previously\cite{SpinDragPRL}.

\textit{Collective modes.---} As a first application of the above, we consider the collective modes. Following the usual
procedure, we linearize the equations and put in a plane-wave ansatz for the hydrodynamical variables. 
Solving for frequency $\omega$ as a function of momentum $k$ results in
\ba
\label{denswave}
\omega = k \sqrt{\frac{1}{m} \frac{\partial p}{\partial n} } \equiv ck, \\
\label{spinwave}
\omega = \frac{A_s(Pn k^2-i\eta_\perp k^4)}{\eta_\perp^2 k^4 + (Pn)^2}
\overset{k\rightarrow0}{\longrightarrow} 
\frac{A_s}{Pn}k^2-\frac{i\eta_\perp A_s}{(Pn)^2}k^4.
\ea
The first equation describes a density wave, first sound, which
propagates at the speed of sound $c$ as expected. The second equation gives the dispersion for the
spin waves and includes their damping. We remark that in the long-wavelength limit it reduces to
a quadratic dispersion with quartic damping, which is in agreement with previous results
for conventional (Fermi) ferromagnets \cite{PhysRev.188.898}. Finally, we did not consider
energy as a hydrodynamic variable and hence neglected the resulting heat diffusion. Thus,
our theory does not properly predict the velocity of second sound.

\begin{figure}[b]
\begin{center}
\includegraphics[width=0.45\linewidth]{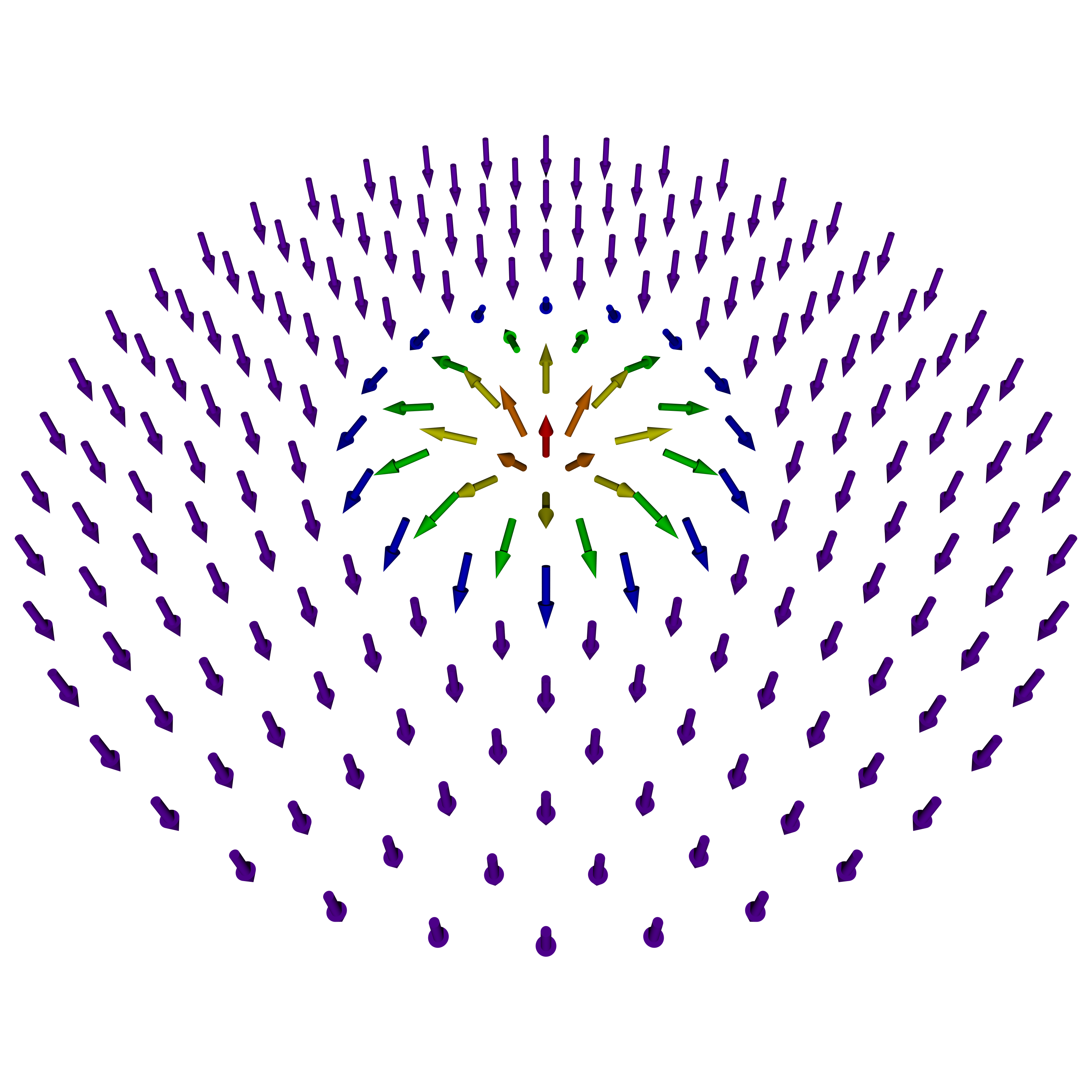}
\includegraphics[width=0.45\linewidth]{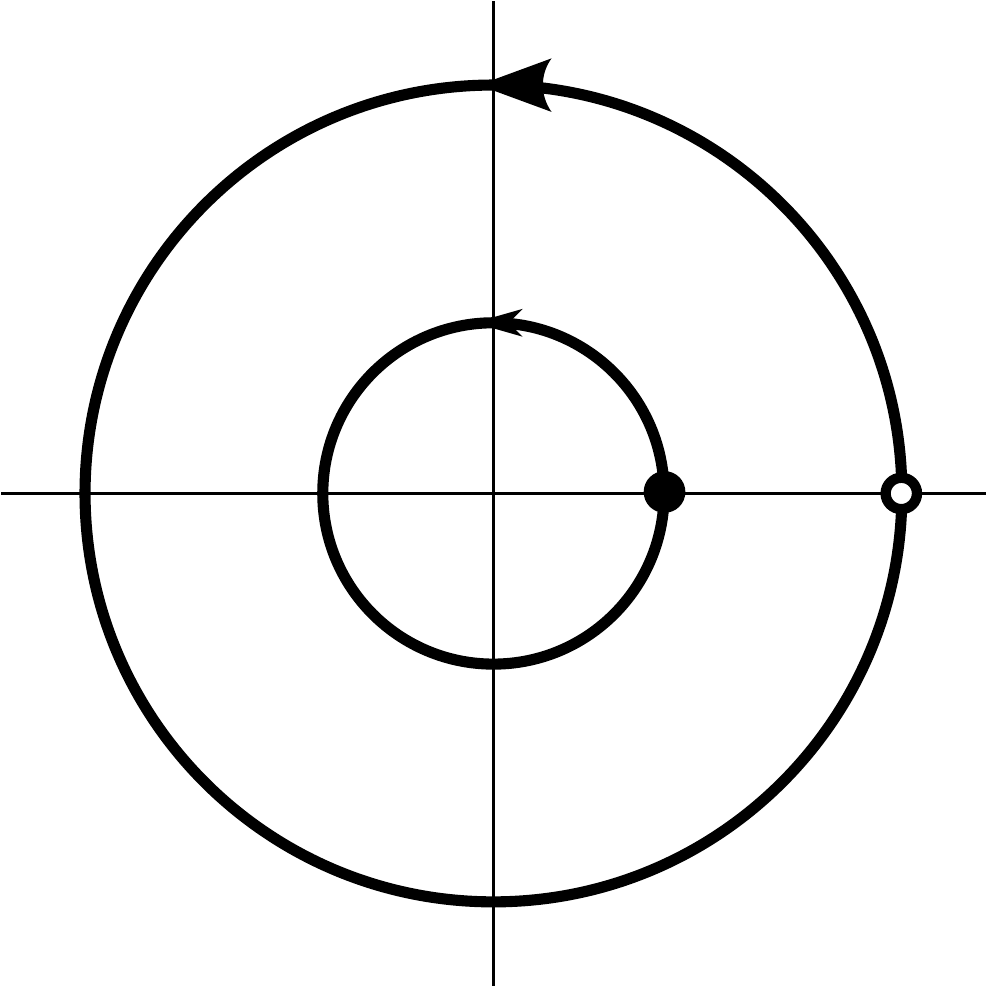}
\caption{A baby skyrmion in a two-dimensional ferromagnetic cloud (\emph{left}). In an isotropic trap, which has
its center in the middle of the picture (\emph{right}), the skyrmion (empty circle) precesses around the center of the trap. The center of the cloud (filled circle) also moves around the center of the trap 
due to the topological Hall effect. The amplitude of the motion of the center of the cloud has been enhanced for clarity.\vspace{-10mm}}
\label{fig:SkyrmionAndCloud}
\end{center}
\end{figure}

\textit{Skyrmion dynamics and topological Hall effect.---} To illustrate the importance of geometrical
forces, we investigate the motion of a 2D or baby skyrmion  (Fig.\ \ref{fig:SkyrmionAndCloud}). 
There are several reasons warranting a closer look at this 2D skyrmion. Firstly, we expect to see
a topological Hall effect, due to the artificial electromagnetic fields generated by the spin texture.
Moreover, due to the interplay between magnetization relaxation and spin gradients, we anticipate 
irreversible dynamics. Lastly, baby skyrmions have been
experimentally realized in spinor Bose gases \cite{PhysRevLett.108.035301}, suggesting that our theory could be confronted
with experiments in the near future.

Considering a rigid skyrmion texture $\oo^\alpha = \oo^\alpha(\bs x - \bs x_\sk)$ with a velocity 
$\bs v_\sk = (\dot x_\sk, \dot y_\sk)$, we derive its equations of motion from our hydrodynamic
description of the dynamics of the atomic cloud. 
To that end, we take a cross product of $\oo^\alpha$ with Eq.\ \eqref{spin}, then take an inner product
with $\bs \nabla \oo^\alpha$ and finally perform an integral over the coordinate space.
Moreover, we assume a pancake-like geometry at a low temperature, and a Thomas-Fermi density profile
for this calculation. 
We define the trap to be isotropic and harmonic in the $xy$-plane with the frequency $\omega_d$.
Furthermore, $\eta_\perp$ is taken to be constant, $A_s(\bs x) = n(\bs x) a_s$,
and $a_s = \hbar P /2m$ as is shown below. 
We also introduce the density in the center of the cloud $n_0$, the density in the
center of the skyrmion $n_\sk$ and the radius of the cloud $R$. The equations of motion ultimately read
\ba
\ddot{x}  + \omega_d^2 x - \Delta \dot y P \hbar I^1_{12} n_\sk / \pi m n_0 R^2 = 0, 
\nonumber \\
\ddot{y} + \omega_d^2 y + \Delta \dot x P \hbar I^1_{12}  n_\sk / \pi m n_0 R^2 = 0,
\nonumber \\
\Delta \dot x (\eta_\perp I^2_{12} + n_\sk P I_{12}^1) 
+ \Delta \dot y  \eta_\perp I^2_{22}
+ \Delta y  n_0 a_s I^3 /R^2 = 0,
\nonumber \\
\Delta \dot y (\eta_\perp I^2_{12} - n_\sk P I_{12}^1) 
+ \Delta \dot x  \eta_\perp I^2_{11}
+ \Delta x n_0 a_s I^3 /R^2 = 0, 
\nonumber
\ea
where $\Delta x = x_\sk - x$ and $\Delta y = y_\sk - y$ are the skyrmion coordinates relative to the center of the cloud.
The quantities $I^1_{ij}$, $I^2_{ij}$ and $I^3$ are real and only depend on the texture
\cite{integrals}, where $i$, $j$ can be $1$ or $2$, corresponding to the $x$ or $y$ direction, respectively.
The integral $I^1_{12}$ is determined by a topological invariant known as the skyrmion
number or the winding number: $W = I^1_{12}/4\pi = 1$. Furthermore, for cylindrically symmetric skyrmion textures, 
$I^3 = h$ and $I^2_{ij} = - \tilde h \delta_{ij} / l^2$, where $h$ and $\tilde h$ are dimensionless
numbers, while $l$ is the length associated with the size of the skyrmion \cite{constants}. The latter integrals are therefore not determined
by the topology of the magnetization texture only.

Following the procedure described above, we find, in addition to the dipole mode with frequency $\omega_d$, 
a collective mode pertaining to the motion of the skyrmion with the frequency
\be
\label{colskyrm}
\omega_\sk =
\frac{h n_0 \hbar \left(4 n_\sk P \pi + i \tilde h \eta_\perp / l^2 \right) }
{2  m R^2 \left(16 n_\sk^2 P^2 \pi^2+\tilde h^2 \eta_\perp^2/ l^4 \right)}.
\ee
The  real part of the frequency implies that the skyrmion is moving around the center of the trap, c.f.\ Fig.\ 
\ref{fig:SkyrmionAndCloud},
while the positive imaginary part means that the skyrmion is spiraling out and is pushed away from the center of the trap. 

We now turn our attention to the case of no damping ($\eta_\perp = 0$).
The eigenvectors of the various modes, written in the form
$(x,y,\Delta x,\Delta  y)$, are $(1,0,0,0)$, $(0,1,0,0)$, $(-i \alpha, \alpha, -i, 1)$ and $(i \alpha, \alpha, i, 1)$ with
\be
\alpha = 
\frac{32 h n_\sk^2 P^2 \pi  \hbar ^2}{64 m^2 n_\sk^2 P^2 \pi ^2 R^4 \omega_d^2-h^2 n_0^2 \hbar ^2}.
\ee
The first two eigenvectors describe the dipole mode, where the cloud 
and the skyrmion move in phase in either the $x$ or the $y$ direction, respectively.
In order to investigate the implications of the third and fourth eigenvectors, we set 
the initial coordinates $\Delta x(0)\neq 0$ and $x(0) = \alpha \Delta x(0)$ with $y(0)=\Delta y(0)=0$. 
We set the initial velocity of the cloud to zero.
In that case, $x$ and $\Delta x$ oscillate in phase with the
frequency $\omega_\sk = h n_0 \hbar/8 m n_\sk P \pi R^2$:
\ba
x = \alpha \Delta x(0) \cos ( \omega_\sk t ), \quad
\Delta x &= \Delta x(0) \cos ( \omega_\sk t ).
\ea
However, both the center of the cloud and the skyrmion start moving in the $y$ direction is as well (cf. Fig. \ref{fig:SkyrmionAndCloud}):
\ba
y &= \alpha \Delta x(0) \left[ 
\sin ( \omega_\sk t ) 
-\frac{\omega_\sk}{\omega_d} \sin( \omega_d t)
 \right],\\
\Delta y &= \Delta x(0) \sin ( \omega_\sk t ),
\ea
due to the force exerted on them by the artificial electromagnetic field, which can also be seen in the equations of motion.
Physically, this effect is a Hall effect as it corresponds to transverse motion in response to a
longitudinal force -- in this case the restoring force of the trapping potential.
Due to the nature of this particular spin texture, this effect is known as the topological Hall effect.

\textit{Microscopic theory.---} We proceed to evaluate the hydrodynamic input parameters $P$, $A_s$ and $\eta_\perp$. To calculate the polarization $P$, we employ the Bogoliubov theory around the ferromagnetic 
groundstate of the gas. That amounts to populating only one of the condensate components, 
$n_c^\ua = n_c \neq 0$, while $n_c^\da =0$.
Scattering amplitudes in ultracold gases are governed by the two-body T matrix $g$ \cite{stoofbook}, which
can have different components for collisions of different spin states. We only consider the case 
with equal T-matrix elements 
$g^{\ua\da} = g^{\ua\ua} = g^{\da\da} = g \neq 0$.
Furthermore, we investigate a balanced mixture, i.e., with equal chemical potentials 
$\mu^\ua = \mu^\da=gn_c^\ua$. This leads to decoupling of the spin components.
The $\ua$ particles obtain the usual Bogoliubov propagator, while the $\da$ particles
retain the non-interacting propagator within this approximation. We thus find the following particle densities in the 
non-condensate states:
\ba
n^{\ua}_\mathrm{nc} = 
\frac{1}{V}
\sum_{\bs k \neq 0}
\left( 
\frac{\ve_k + gn_c}{\hbar \omega_k}
\right. &
\frac{1}{e^{\beta\hbar\omega_k}-1} 
\nonumber \\
&
+
\left.
\frac{\ve_k + gn_c - \hbar \omega_k}{2\hbar \omega_k}
\right),
\ea
and $n^{\da}_{nc} =
\sum_{\bs k \neq 0}
[\exp({\beta\ve_k})-1]^{-1}/V,$
where the Bogoliubov dispersion $\hbar \omega_k = \sqrt{\ve_k (\ve_k +2 gn_c)}$, $\beta=1/k_\mathrm{B} T$ is the inverse thermal
energy
and $\ve_k = \hbar^2 k^2/2m$ is the free particle dispersion. The density distributions are 
subject to the constraint
$n = n_c + n^\ua_{nc} + n^\da_{nc}$.
We note that after fixing the total density $n$, temperature and interaction
strength, we can solve these equations for $n_c$ and obtain $n_{nc}^\ua$ as well
as $n_{nc}^\da$ at the same  time. This gives us the polarization of the gas
\be
P = (n_c + n^{\ua}_{nc} - n^{\da}_{nc})/n
=1 - 2n^{\da}_{nc}/n
.
\ee

In the intermediate temperature regime $ ng \ll k_B T < k_B T_\text{BEC}$ it is straightforward to calculate $P$ analytically.
To that end, we approximate $\hbar \omega_k \simeq \ve_k + gn_c$, so that
$
n_{nc}^\ua=n_{nc}^\da = \zeta(3/2) (mk_\mathrm{B} T/2\pi\hbar^2)^{3/2}
$
and
\be
P = 1 - 2 \zeta(3/2) (mk_B T/2\pi\hbar^2)^{3/2}/n,
\ee
where $\zeta$ denotes the Riemann zeta function. Note that in this approximation we have 
$P = 1$ at zero temperature, as the depletion of the condensate is neglected and all the ideal gas particles end up in the condensate.

We now turn to the calculation of the spin stiffness $A_s$. 
According to the Bogoliubov theory,
when only one spin component is populated with a Bose-Einstein condensate, excitations in the other spin components
are free particles, corresponding to spin waves. Therefore, the dispersion of the spin waves is simply
$\hbar \omega_k = \ve_k$. Comparing this with the real part of the dispersion relation for the spin
waves given by the hydrodynamics in Eq.\ \eqref{spinwave}, we conclude that
\be
\label{eq:stiffness}
A_s = \frac{\hbar}{2m} P n.
\ee

The only quantity that remains to be evaluated is $\eta_\perp$.
An upper bound for this quantity is found by considering the non-condensed phase.
In this case, we note that it is equal to the transverse spin conductivity \cite{PhysRevB.79.094415}
\be
\eta_\perp = \hbar \sigma_\perp.
\ee
We calculate $\sigma_\perp$ in the normal phase 
from a set of Boltzmann equations and use the relaxation-time approximation. 
For a given species, e.g.\ $\ua$, the momentum-dependent 
relaxation time $(\tau^{\ua\ua})_k$ is given by the well-known collision integral
\ba
\frac{1}{(\tau^{\ua\ua})_k} =
&\frac{2\pi}{\hbar} g_{\ua\da}^2
\int 
\frac{\dd \bvk_2}{(2\pi)^3}
\frac{\dd \bvk_3 }{(2\pi)^3}
\frac{\dd \bvk_4}{(2\pi)^3}
f^{\da\da}(\bvk_2) [1+ f^{\ua\ua}(\bvk_3)]
\nonumber \\
&\times
[1+f^{\da\da}(\bvk_4)]
(2\pi)^3\delta^{(3)}(\bvk+\bvk_2-\bvk_3-\bvk_4)
\nonumber \\
&\times \delta(\ve_k + \ve_{k_2}-\ve_{k_3}-\ve_{k_4}),
\ea
that can be derived
using Fermi's golden rule. Here we only consider the term due to inter-species scattering, as
intra-species scattering terms drive the distributions $f^{\ua\ua}$ and $f^{\da\da}$ towards the Bose-Einstein
equilibrium distribution.

\begin{figure}[h]
\begin{center}
\includegraphics[width=0.9\linewidth]{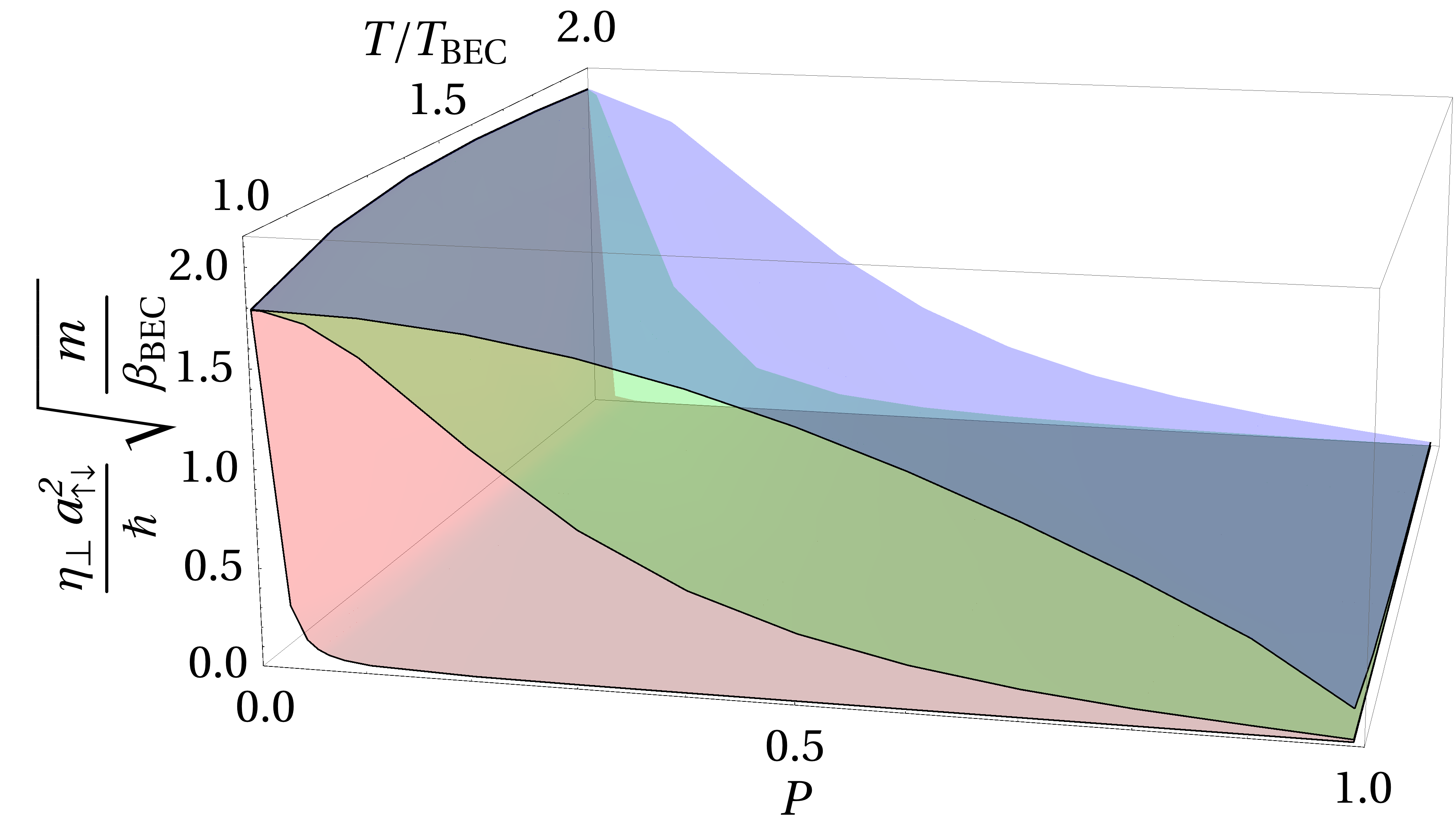}
\caption{Transverse spin diffusion $\eta_\perp$ as a function of polarization $P$ and temperature
for a homogeneous gas. The three surfaces correspond to $n g_{\ua\da} \beta_\text{BEC} = 1$, $0.5$ and $0.1$ from
top to bottom. Here $a_{\ua\da}$ is the inter-species scattering length and $\beta_\text{BEC} \equiv 1/k_B T_\text{BEC}$.\vspace{-8mm}}
\label{fig:sigma}
\end{center}
\end{figure}

The transverse spin relaxation time is then given by
\be
2/(\tau^\perp_s)_k = 1/(\tau^{\ua\ua})_k + 1/(\tau^{\da\da})_k.
\ee
Given this relaxation time, the transverse spin conductivity is obtained by evaluating the 
integral
\be
\eta_\perp
=
-
\frac{\hbar^3}{m^2}
\int 
\dd \bvk
k^2
\frac{(\tau_s^\perp)_k}{1
+[(\tau_s^\perp)_k \Delta/\hbar]^2} 
\frac{\partial f_0}{\partial \ve_k},
\ee
where $f_0$ is the sum of the equilibrium distributions of $\ua$ and $\da$ particles and 
$\Delta=\mu^\ua - \mu^\da$ is the exchange splitting. For a particular kind of atom, $\eta_\perp$ depends
on the temperature and all the scattering lengths in the
system. However, by fixing the inter-species T matrix
$g_{\ua\da}$, we can obtain a more general picture as a function
of temperature and polarization as
shown in Fig. \ref{fig:sigma}.

\textit{Discussion and conclusion.---} In summary, we have constructed a hydrodynamic theory for a
ferromagnetic spin-1/2 Bose gas that is valid for any temperature relevant for cold-atom systems. We have also calculated all the
input parameters for the hydrodynamic theory within the Bogoliubov approximation. Finally, we have considered dynamics of a topological spin texture
(skyrmion) and found it to lead to a topological Hall effect. 

To determine if the topological Hall effect can be
observed experimentally, we estimate the skyrmion precession frequency $\omega_\sk$ and
the eigenvector parameter $\alpha$. Considering a condensate of $10^4$ $^{23}$Na atoms in a pancake-like geometry with radial confinement $\omega_\perp/2\pi = 1\, \text{Hz}$ and perpendicular confinement $\omega_z = 10 \omega_\perp$,
we estimate $\omega_\sk = \omega_\perp$ and $\alpha = 0.1$. This corresponds to a cloud size
of $50\, \mu m$ and a Hall amplitude of $5\, \mu m$, which should be observable with current
experimental techniques.

When it comes to the damping of spin waves, we consider a homogeneous $^{87}$Rb gas ($\vert F = 1, m_f = -1 \rangle$ and
$\vert F = 2, m_f =1 \rangle$) with 
the density of $10^{16}$ cm$^{-3}$ as an example.
In particular, at the Bose-Einstein condensation 
temperature and polarization $P=1/2$ a spin wave with momentum $k = 1 \, \mu m^{-1}$, which is within 
the reach of current experiments, has a damping time of $0.7$ s.

In the future work, we plan to calculate $\eta_\perp$ in the superfluid phase ($T < T_\text{BEC}$),
which would complete the microscopic input for the parameters of the hydrodynamic theory in the
entire temperature range.
Moreover, it is worthwhile to apply the current approach to higher spin systems, where
extra degrees of freedom such as the nematic tensor \cite{PhysRevA.86.063614} enter the theory.

This work is supported by the Stichting voor Fundamenteel Onderzoek der Materie (FOM) and the Nederlandse Organisatie voor Wetenschaplijk Onderzoek (NWO).
\vspace{-5mm}

\bibliography{prl1}

\begin{thebibliography}{38}
\bibitem [{con()}]{constants}%
  \BibitemOpen
  \href@noop {} {}\bibinfo {note} {For a magnetization texture $\oo^\alpha =
  (\sin A(r) cos \phi, \sin A(r)\sin\phi, \cos A(r))$, where $A(r) = 2 \arctan
  (r/l)$, on a plane parameterized by polar coordinates $(r, \phi)$, we find $h
  = 8\pi$, $\tilde h = 16\pi/3$.}\BibitemShut {Stop}%
\end{thebibliography}

\end{document}